\documentclass[letterpaper,twocolumn]{article}

\usepackage{amsmath,amssymb,amsthm}
\usepackage{graphicx}
\usepackage{hyperref}
\usepackage{algorithm}
\usepackage{algpseudocode}
\usepackage{booktabs}
\usepackage{xcolor}
\usepackage{siunitx}
\usepackage[margin=1in]{geometry}

\allowdisplaybreaks 

\DeclareMathOperator{\diag}{diag}
\DeclareMathOperator*{\argmax}{argmax}

\newtheorem{remark}{Remark}

\begin{document}

\title{Real-Time Threat Classification in Multi-Agent Systems via Topology-Coupled Hidden Markov Models}

\author{
  Ailin Barzegar, Davide Spinello, and Mohammad Pirani\\
  Department of Mechanical Engineering, University of Ottawa, Ottawa, ON, Canada
}

\maketitle

\begin{abstract}
Networked multi-agent systems coordinating in adversarial environments face compound cyber threats that alternate between false-data injection (FDI), denial-of-service (DoS), and stealthy decoy perturbations designed to mimic benign operational anomalies. Because the switching policy is latent and the plant dynamics evolve continuously, deterministic threshold detectors struggle to isolate deception without causing excessive false alarms. This paper presents an online, data-driven framework that couples distributed continuous time unknown input observers (UIOs) with a topology-aware Hidden Markov Model (HMM) for real-time attack detection and classification under unknown adversarial strategies. By decoupling FDI signals from augmented node states, distributed observers deployed across the network generate local augmented residuals that are discretized into emission sequences. The HMM embeds the underlying directed communication graph into the off-diagonal blocks of its transition probability matrix, explicitly capturing inter-agent attack propagation. Both transition and observation probabilities are updated recursively online via an Expectation-Maximization (EM) scheme, adapting the classifier to evolving threat behaviours without prior offline calibration. Numerical simulations on a directed network demonstrate that incorporating graph topology into the latent transition dynamics resolves the severe classification ambiguity between nominal and decoy modes, outperforming decoupled agent-independent decoders while restricting observer deployment to a sparse subset of nodes.
\end{abstract}

{{\bf keywords:} Cyber-attack classification; Multi-agent systems; Resilient cooperative control; Hidden Markov models; Unknown input observers; Markovian jump systems}

\begin{center}
\footnotetext[1]{Preprint under review in the ASME Journal of Dynamic Systems, Measurement, and Control.}
\end{center}

\section{Introduction}
The growth of networked cyber-physical systems has placed attack resilient control of networked systems at the forefront of control engineering research \cite{11175202,xing2024,wang2023resilient,segovia-ferreiraSurveyCyberResilienceApproaches2024}. In multi-agent systems (MASs), distributed agents coordinate through shared communication links to accomplish collective objectives ranging from formation flight and sensor fusion to smart grid regulation~\cite{xing2024,zhang2026observer}. The inter-agent coupling, while essential for performance, is also a structural liability: the compromise of a single node can cascade harmful effects across the network, and the simultaneity of threats arriving at multiple locations compounds the difficulty of both detection and response~\cite{xu2025}.

From an attacker's perspective, the most tactically effective strategies are those that blend into normal operations; therefore, they rarely rely on a single, persistent attack type. Instead, they alternate between false-data injection (FDI), which corrupts state or measurement signals, and denial-of-service (DoS), which disrupts communication links, while also deploying decoy perturbations that mimic anomalous patterns to trigger false alarms and saturate the defender's detection resources~\cite{yang2024,kushwaha2026stealthy}. This compound threat landscape poses a harder problem than the isolated attack scenarios that are extensively addressed in the existing literature~\cite{xu2019}.

In tactical multi-agent networks, such as those governing autonomous vehicle countermeasures and operations in spectrum-denied environments, it is crucial to evaluate model-based attack detection and mitigation. While existing literature has explored unknown input observers (UIOs) for deterministic cybersecurity and fault tolerance, including multi-observer estimation architectures \cite{yang2020unknown} and distributed state reconstruction \cite{cao2023distributed}, these approaches often struggle against stochastic decoy perturbations engineered to mimic benign anomalies. Consequently, robust resilient control requires fusing deterministic decoupled residuals with data-driven sequential classifiers to resolve mode ambiguity.

The vulnerability of deterministic observers to sophisticated decoy perturbations is rooted in the algebraic limitations of cyber-physical attack detectability. As established in \cite{pasqualetti2013attack}, an adversary with sufficient knowledge of the system's dynamics can design ``stealthy'' or zero-dynamics attacks that are bounded signals that hide entirely within the system's null space. Because these zero-dynamics attacks generate output residuals that are geometrically indistinguishable from nominal operation or localized benign anomalies \cite{pasqualetti2012consensus}, static UIOs and standard threshold monitors are easily saturated or bypassed. Resolving the ambiguity of these stealthy decoy modes therefore requires augmenting the deterministic residual generation with a data-driven, probabilistic framework capable of learning the adversary's stochastic switching behaviour over time.

Modeling such behaviour, particularly in the absence of information about attacker strategies, requires moving beyond deterministic attack representations. In this work, the physical dynamics of each agent are described in continuous time, while the attack mode switches discretely in time, forming a Markovian jump system. This separation reflects the physical reality that the plant evolves continuously, but the adversarial mode changes are instantaneous events. Stochastic frameworks and Markovian jump processes in particular are well-suited for capturing the intermittent and partially unknown nature of adaptive adversaries~\cite{shi2015,hedayati2025}. A key complication, however, is that the Markovian mode sequence is generally unknown or not directly observed by the individual agents: the defender perceives only output residuals, which conflate the contributions of different attack types and are further obscured by decoy signals. Existing Markov-based security schemes sidestep this difficulty by assuming mode observability~\cite{chen1995,durrett1999}, an assumption that becomes invalid when the attacker's strategies are almost unknown.

Hidden Markov Models (HMMs) allow for performing inference over such latent mode sequences from observable residual data~\cite{rabiner1989tutorial}. Prior HMM-based approaches in network security have demonstrated their value for sequential attack recognition~\cite{chadza2020}, but they typically treat the model parameters as static and pre-calibrated, and they do not incorporate the communication graph structure into the transition dynamics. As a result, they cannot capture how an attack originating at one agent may probabilistically trigger mode transitions in its neighbours, a propagation effect that is central to multi-hop adversarial strategies~\cite{ding2024}.

In this context, this paper makes three specific contributions:

\begin{enumerate}
  \item We introduce a graph-structured stochastic threat model in which each agent's mode evolves over four states named nominal, FDI, DoS, and decoy via a Markov chain whose inter-agent transition blocks are weighted by the adjacency matrix of the communication graph. This allows the model to represent cross-network attack propagation without requiring a separate propagation layer.

  \item We construct an online HMM classifier that pairs a distributed UIO with an iterative Expectation-Maximization (EM) update procedure. The Observation Probability Matrix (OPM) of the HMM is parameterized by Gaussian likelihoods estimated from historical residuals, and both the Transition Probability Matrix (TPM) and OPM are corrected as new measurements become available, enabling classification under evolving, partially known adversarial behaviour.

  \item We demonstrate on a directed network of seven agents that the proposed framework achieves classification accuracy significantly above the random baseline across all four modes, including the statistically challenging decoy mode, and we report per-mode precision, recall, and F1-score to facilitate comparison with future methods.
\end{enumerate}

The rest of the paper is organized as follows. Section~\ref{sec:problem} develops the system model and threat formulation. Section~\ref{sec:method} details the observer design, residual discretization, and HMM construction. Section~\ref{sec:sim} presents simulation results. Section~\ref{sec:conclusion} closes with limitations and research directions.

\section{Problem Formulation}
\label{sec:problem}

\subsection{Agent Dynamics Under Markovian Mode Switching}

Consider a system composed of $M$ agents whose interactions are encoded in a weighted directed graph $\mathcal{G} = (\mathcal{V}, \mathcal{E})$, with vertex set $\mathcal{V} = \{1,\ldots,M\}$, edge set $\mathcal{E} \subseteq \mathcal{V} \times \mathcal{V}$, and non-negative adjacency weight $w_{jm}$ assigned to each link $(j,m) \in \mathcal{E}$.

At a given time instant, the input to each agent is in one of four operational modes drawn from the set $\Theta \triangleq \{\mathrm{N}, \mathrm{FDI}, \mathrm{DoS}, \mathrm{D}\}$, representing {\it nominal operation, false-data injection, denial-of-service,} and {\it decoy behaviour}, respectively. The physical plant of each agent evolves in continuous time, while the mode index switches at discrete instants $k = 0, 1, 2,\ldots$. For agent $m$, mode transitions across all four states are described by a discrete-time Markovian process with state $\theta_m \in \Theta$ that is not accessible to the defender, inducing a hidden state inference problem. Therefore, the mode sequences $\{\theta_m(k)\}_{k \geq 0}$, $m=1\ldots,M$, are latent in the sense that no agent can directly observe the current mode.

The continuous time dynamics of the $m$-th agent are
\begin{subequations}\label{eq:agent_dyn}
\begin{align}
\dot{x}_m(t) &= a_m x_m(t) + b_m u_m(t,\theta_m(k)) \nonumber \\
           &\quad + f_m \psi_m(t,\theta_m(k)) + d_m \xi_m(x_m(t),\theta_m(k)),\\
y_m(t)     &= c_m x_m(t),
\end{align}
\end{subequations}
where $x_m \in \mathbb{R}^{n_x}$ is the state, $u_m \in \mathbb{R}^{n_u}$ is the control input, $\psi_m \in \mathbb{R}^{n_\psi}$ is the FDI injection signal (active only in FDI mode), $\xi_m \in \mathbb{R}^{n_\xi}$ is the mode-dependent decoy perturbation, and $y_m \in \mathbb{R}^{n_y}$ is the $m$-th agent's output, which is the node's observable. The integers $n_x$, $n_u$, $n_\psi$, $n_\xi$, and $n_y$ denote the dimensions of the corresponding signals. The system matrices $a_m,b_m,f_m,d_m,c_m$ have compatible dimensions and are time-invariant.

In nominal operating conditions, the control input $u_m$ encodes the standard consensus protocol via the underlying network topology~\cite{olfati2004consensus}. Let $\mathcal{N}_m \subset \mathcal{V}$ be the set of nodes (neighbours) connected to node $m$, that is, the set of nodes for which a directed edge exists with node $m$. In continuous time, the nominal consensus control is:
\begin{equation}\label{eq:u_N}
    u_m(t, \theta_m(k) = \mathrm{N}) = \sum_{j \in \mathcal{N}_m} w_{jm} \left(x_j(t) - x_m(t)\right).
\end{equation}
By defining the weighted $M \times M$ adjacency matrix $\mathcal{A}$ with entries $[\mathcal{A}]_{mj} = w_{jm}$, and the diagonal in-degree matrix $\mathcal{D}$ with non-zero entries $[\mathcal{D}]_{mm} = \sum_{j=1}^{M} w_{jm}$, the digraph Laplacian is given by $\mathcal{L} = \mathcal{D} - \mathcal{A}$, and the consensus control input for the $m$-th agent in nominal mode in continuous time can be written as 
\begin{subequations}
\begin{align}
u_m\bigl(t,\theta_m(k)=\mathrm{N}\bigr) &= -\sum_{j=1}^{M} [\mathcal{L}]_{mj} x_j(t), \\
\psi_m\bigl(t,\theta_m(k)=\mathrm{N}\bigr) &= 0, \\
\xi_m\bigl(t,\theta_m(k)=\mathrm{N}\bigr) &= 0.
\end{align}
\end{subequations}
The DoS mode is modelled as an intermittent loss of control input $u_m(t)$, reflecting packet-drop attacks on the communication channel:
\begin{subequations}
\begin{align}
u_m(t, \theta_m(k) = \mathrm{DoS}) &= 0, \\
\psi_m\bigl(t,\theta_m(k)=\mathrm{DoS}\bigr) &= 0, \\
\xi_m\bigl(t,\theta_m(k)=\mathrm{DoS}\bigr) &= 0.
\end{align}
\end{subequations}
When the attack is in FDI mode, the control $u_m$ is still given by \eqref{eq:u_N}, but with an additive non-zero input signal $\psi_m(t,\theta_m(k))$:
\begin{subequations}
\begin{align}
u_m\bigl(t,\theta_m(k)=\mathrm{FDI}\bigr) &= -\sum_{j=1}^{M} [\mathcal{L}]_{mj} x_j(t), \\
\psi_m\bigl(t,\theta_m(k)=\mathrm{FDI}\bigr) &= \text{Additive fault signal}, \\
\xi_m\bigl(t,\theta_m(k)=\mathrm{FDI}\bigr) &= 0.
\end{align}
\end{subequations}

Among the latent modes, the decoy is the most challenging in terms of detection, since the attacker strategy is to craft a Lipschitz-bounded signal whose statistical footprint overlaps with other operational modes, causing threshold-based detectors to either miss the threat or issue false alarms on benign data. The decoy mode function $\xi_m$ is therefore assumed to satisfy the Lipschitz condition \cite{rahimi2022neighborhood}:
\begin{equation}
\left\| \xi_m\big(x_{m1}(t)\big) - \xi_m\big(x_{m2}(t)\big) \right\| \leq L_{m} \left\| x_{m1}(t) - x_{m2}(t) \right\|
\label{eq:lipschitz}
\end{equation}
for a known constant $L_m > 0$, producing residuals that overlap with those of other modes, making mode classification challenging:
\begin{subequations}
\begin{align}
u_m\bigl(t,\theta_m(k)=\mathrm{D}\bigr) &= -\sum_{j=1}^{M} [\mathcal{L}]_{mj} x_j(t), \\
\psi_m\bigl(t,\theta_m(k)=\mathrm{D}\bigr) &= 0, \\
\xi_m\bigl(t,\theta_m(k)=\mathrm{D}\bigr) &= \text{Additive decoy signal}.
\end{align}
\end{subequations}


The latent mode dynamics of the multi-agent Markov process are described by the intra-agent transition probabilities and the cross-agent conditional probabilities for all modes $p, q \in \Theta$, focal agents $m$ and neighbour agents $j$:
\begin{subequations}\label{eq:tpm_true}
\begin{align}
T_{pq}^{mm} &= P(\theta_{m}(k+1)=q \mid \theta_{m}(k)=p), \\
T_{pq}^{mj} &= P(\theta_{m}(k+1)=q \mid \theta_{j}(k)=p),
\end{align}
\end{subequations}
where $\sum_{q} T_{pq}^{mm} = 1$ and $\sum_{q} T_{pq}^{mj} = 1$. Since $\theta_m(k)$ is unobservable, the defenders estimate their current mode via residuals produced by the distributed observers introduced in Section~\ref{sec:method}, whose statistical relationship to the hidden modes is captured by the OPM discussed in Section~\ref{sec:opm}.

\begin{remark} 
The index $k$ in \eqref{eq:tpm_true} advances at a fixed period chosen to be equal to the observer sampling period \(h\) introduced in Section~\ref{sec:residual}. This ensures that each HMM observation window is aligned with exactly one potential mode transition interval.
\end{remark}

\subsection{Distributed Augmented Model}

To monitor the network while respecting communication constraints, a distributed estimation architecture is employed where each agent is equipped with a local observer. By fusing measurements along cooperative paths within its communication neighbourhood, each observer reconstructs the augmented state of the agent and its immediate neighbours.


To formalize this, let $\mathcal{N}_m = \{j_1,\dots,j_{|\mathcal{N}_m|}\}$ denote the set of neighbours of agent $m$ with cardinality $|\mathcal{N}_m|$. Define the set of indices $\{m \}\cup \mathcal{N}_m$ composed of agent $m$ and its neighbours, and define the augmented state for agent $m$ as the stack
\begin{equation}
X_m(t)=\big[x_m^\top(t),\,x_{j_1}^\top(t),\,\dots,\,x_{j_{|\mathcal{N}_m|}}^\top(t)\big]^\top\in\mathbb{R}^{n_x(1+|\mathcal{N}_m|)},
\end{equation}
where $X_m(t)$ is the stacked state vector of agent $m$ and its neighbours. Define the block-diagonal matrices
\begin{subequations}
\begin{align}
A_m &= \diag\left(a_m, a_{j_1}, \dots, a_{j_{|\mathcal{N}_m|}}\right),\\
B_m &= \diag\left(b_m, b_{j_1}, \dots, b_{j_{|\mathcal{N}_m|}}\right),\\
C_m &= \diag\left(c_m, c_{j_1}, \dots, c_{j_{|\mathcal{N}_m|}}\right),\\
D_m &= \diag\left(d_m, d_{j_1}, \dots, d_{j_{|\mathcal{N}_m|}}\right),\\
F_m &= \diag\left(f_m, f_{j_1}, \dots, f_{j_{|\mathcal{N}_m|}}\right),
\end{align}
\end{subequations}
where the blocks of agent $m$ are listed first, followed by those of its neighbours in the same order as in $\mathcal{N}_m$. To map the single-agent signals to the augmented model, let $\boldsymbol{\theta}(k) = [\theta_1(k), \dots, \theta_M(k)]$ denote the joint mode vector of the network, and define the stacked control input, unknown fault input, and decoy vectors as:
\begin{subequations}\label{eq:aug_vectors}
\begin{align}
& U_m(t) = \big[u_m^\top,\, u_{j_1}^\top, \ldots, u_{j_{|\mathcal{N}_m|}}^\top \big]^\top, \\
& \Psi_m(t, \boldsymbol{\theta}(k)) = \big[\psi_m^\top(t, \theta_m(k)), \psi_{j_1}^\top(t, \theta_{j_1}(k)), \ldots \nonumber \\ 
    & \quad \ldots \psi_{j_{|\mathcal{N}_m|}}^\top(t, \theta_{j_{|\mathcal{N}_m|}}(k))\big]^\top, \\
& \Xi_m(X_m(t), \boldsymbol{\theta}(k)) = \big[\xi_m^\top(x_m(t), \theta_m(k)), \xi_{j_1}^\top(x_{j_1}(t), \theta_{j_1}(k)), \dots \nonumber \\
& \quad \ldots \xi_{j_{|\mathcal{N}_m|}}^\top(x_{j_{|\mathcal{N}_m|}}(t), \theta_{j_{|\mathcal{N}_m|}}(k))\big]^\top.
\end{align}
\end{subequations}
Then, the continuous time dynamics of the augmented states are written as
\begin{subequations}\label{eq:aug_dyn}
\begin{align}
  \dot{X}_m(t) &= A_m X_m(t) + B_m U_m(t,\boldsymbol{\theta}(k)) + F_m\Psi_m(t,\boldsymbol{\theta}(k)) \nonumber \\
               &\quad + D_m\Xi_m(X_m(t),\boldsymbol{\theta}(k)), \\
  Y_m(t)       &= C_m X_m(t),
\end{align}
\end{subequations}
where the individual consensus control input $u_m(t)$ follows the protocol defined in~\eqref{eq:u_N}. Network-level consensus requires:
\begin{equation}
  \lim_{t \to \infty} \|x_i(t) - x_j(t)\| = 0, \quad \forall\, i,j \in \{1,\ldots,M\}.
\end{equation}

\section{Data-Driven Attack Classification via Hidden Markov Models}
\label{sec:method}

\subsection{Residual Generation}
\label{sec:residual}

Adversarial FDI signals act as unknown inputs, so a UIO is used to decouple them from the state estimate, keeping the residuals sensitive to mode changes rather than to the attack signal itself~\cite{barzegar2023}. The observer (Appendix~A) reconstructs the augmented state $\hat X_m(t)$, giving the residuals
\begin{subequations}
\begin{align}
e_m(t) &= X_m(t)-\hat X_m(t),\\
r_m(t) &= Y_m(t)-\hat Y_m(t).
\end{align}
\end{subequations}
These continuous time signals are sampled at a period $h>0$ to obtain $r_m(k)$, which serves as the observable linking the physical system to the hidden modes.

\subsection{Residual Discretization via Mode-Dependent Thresholds}

Before building the HMM, the continuous residual stream must be converted into a discrete one suitable for probabilistic sequence modeling. This is accomplished through an offline threshold tuning stage followed by an online binning rule.

In offline threshold tuning, the augmented system is simulated under each of the four modes in isolation, with the mode identity known throughout. The discretized residual norm $\|r_m(k)\|$ is logged for every agent $m$ over a sufficiently long horizon, and the sample mean $\mu_{m,\theta}$ and standard deviation $\sigma_{m,\theta}$ are extracted for each process $\theta_m$. These statistics define the interval boundaries~\cite{komadina2024}:
\begin{subequations}\label{eq:thresholds}
\begin{align}
  \tau^{\text{low}}_{m,\theta}  &= \mu_{m,\theta} - \kappa_\sigma \sigma_{m,\theta}, \\
  \tau^{\text{high}}_{m,\theta} &= \mu_{m,\theta} + \kappa_\sigma \sigma_{m,\theta},
\end{align}
\end{subequations}
where $\kappa_\sigma > 0$ is a chosen multiplier; $\kappa_\sigma = 3$ is standard practice, placing the bounds at three standard deviations from the mode's mean. Because the bounds are derived separately for each agent and each mode, they automatically account for heterogeneous attack magnitudes and local topological effects.

For online binning, at each time step $k$, the norm $\|r_m(k)\|$ is compared against the stored intervals and assigned to one of four discrete symbols $b_m(k) \in \{0,1,2,3\}$, where $0=\mathrm{N}$, $1=\mathrm{FDI}$, $2=\mathrm{DoS}$, and $3=\mathrm{D}$. Combining observations across agents gives the observation vector $\boldsymbol{b}(k) = [b_1(k), \ldots, b_M(k)]$, which is fed to the HMM at every step.

\subsection{Hidden Markov Model Setup for Multi-Agent Systems}

Let the joint HMM state space be defined by coupling the agents to their respective operational modes:
\begin{equation}
  \mathcal{S} = \bigcup_{m=1}^M \{ (m, \theta) \mid \theta \in \Theta \},
\end{equation}
yielding $N_{\mathcal{S}} = 4M$ total states. The HMM is parameterized by~\cite{chadza2020}
\begin{equation}
  \lambda = (\boldsymbol{T},\, \boldsymbol{O},\, \pi),
\end{equation}
where $\boldsymbol{T} \in \mathbb{R}^{N_{\mathcal{S}} \times N_{\mathcal{S}}}$ is the TPM, $\boldsymbol{O} \in \mathbb{R}^{N_{\mathcal{S}} \times 4}$ is the OPM, and $\pi \in \mathbb{R}^{N_{\mathcal{S}}}$ is the initial mode distribution.

\subsection{Characterization of the Observation Probability Matrix} \label{sec:opm}

Each entry of the OPM represents the probability of observing the residual bin $b$ given that agent $m$ is in mode $\theta$:
\begin{equation}
  P_{(m,\theta),b} = P\left(\text{residual in bin } b \mid (m,\theta)\right).
\end{equation}
This is computed as the normalized Gaussian likelihood at the bin centre $c_b$, where $\phi(x; \mu, \sigma^2)$ denotes the probability density function of a normal distribution with mean $\mu$ and variance $\sigma^2$~\cite{jung2022}:
\begin{equation}
  P_{(m,\theta),b} =
    \frac{\phi(c_b;\, \mu_{m,\theta},\, \sigma^2_{m,\theta})}
    {\sum_{j=0}^{3} \phi(c_j;\, \mu_{m,\theta},\, \sigma^2_{m,\theta})},
  \label{eq:popm}
\end{equation}
where $\{\mu_{m,\theta}, \sigma^2_{m,\theta}\}$ is estimated from historical residuals. Normalization guarantees a valid probability distribution. The resulting OPM is:
\begin{subequations}
\begin{align}
&\boldsymbol{O} = \begin{bmatrix}
    O^{(1)} & O^{(2)} & \cdots & O^{(M)}
  \end{bmatrix}^\top \in\mathbb{R}^{4M\times 4},
  \\
&O^{(m)} = \begin{bmatrix}
    P_{(m,\mathrm{N}),0}   & P_{(m,\mathrm{N}),1}   & P_{(m,\mathrm{N}),2}   & P_{(m,\mathrm{N}),3} \\
    P_{(m,\mathrm{FDI}),0} & P_{(m,\mathrm{FDI}),1} & P_{(m,\mathrm{FDI}),2} & P_{(m,\mathrm{FDI}),3} \\
    P_{(m,\mathrm{DoS}),0} & P_{(m,\mathrm{DoS}),1} & P_{(m,\mathrm{DoS}),2} & P_{(m,\mathrm{DoS}),3} \\
    P_{(m,\mathrm{D}),0}   & P_{(m,\mathrm{D}),1}   & P_{(m,\mathrm{D}),2}   & P_{(m,\mathrm{D}),3}
  \end{bmatrix}.
\label{eq:opm}
\end{align}
\end{subequations}

\subsection{Topology-Aware Transition Probability Matrix}
\label{sec:tpm}

The TPM captures both intra-agent mode transitions and the influence of neighbouring agents through the communication network. For $M$ agents, the unnormalized transition matrix is structured as:
\begin{equation}
  \tilde{\boldsymbol{T}} = \begin{bmatrix}
    P_{11}      & w_{12}P_{12} & \cdots & w_{1M}P_{1M} \\
    w_{21}P_{21} & P_{22}     & \cdots & w_{2M}P_{2M} \\
    \vdots   & \vdots       & \ddots & \vdots \\
    w_{M1}P_{M1} & w_{M2}P_{M2} & \cdots & P_{MM}
  \end{bmatrix},
  \label{eq:tpm}
\end{equation}
%
where the diagonal blocks $P_{mm} \in \mathbb{R}^{4 \times 4}$ contain the estimated intra-agent transition probabilities approximating the true latent dynamics $T_{pq}^{mm}$ defined in~\eqref{eq:tpm_true}, and the off-diagonal blocks $P_{mj} \in \mathbb{R}^{4 \times 4}$ contain the estimated cross-agent conditional probabilities approximating $T_{pq}^{mj}$, scaled by the adjacency weight $w_{jm}$~\cite{pernes2019}. To ensure the global matrix $\boldsymbol{T}$ is a valid row-stochastic transition probability matrix, it is obtained from $\tilde{\boldsymbol{T}}$ via row normalization:
\begin{equation}\label{eq:Tnormalization}
    \boldsymbol{T} = \Lambda^{-1} \tilde{\boldsymbol{T}},
\end{equation}
where $\Lambda$ is a diagonal matrix containing the row sums of $\tilde{\boldsymbol{T}}$.


Each diagonal block $P_{mm}$ is populated from empirical mode-transition counts extracted from the training sequence. The off-diagonal blocks $P_{mj}$ reflect joint transition frequencies between agents $m$ and $j$ over the same sequence, subsequently scaled by the adjacency weight $w_{jm}$. Both sets of blocks are refined by the EM algorithm described in Section~\ref{sec:EM}.

\subsection{Online Expectation-Maximization-Based Parameter Update}
\label{sec:EM}

The HMM parameters are updated recursively to maximize the log-likelihood of the observation sequence $\boldsymbol{b}_{1:\tau} = \{\boldsymbol{b}(1), \ldots, \boldsymbol{b}(\tau)\}$ via the Expectation-Maximization (EM) framework~\cite{dempster1977maximum, baum1970maximization}:
\begin{equation}
  \ell(\lambda) = \log P(\boldsymbol{b}_{1:\tau} \mid \lambda).
  \label{eq:likelihood}
\end{equation}



E-step: Using the forward-backward algorithm~\cite{rabiner1989tutorial}, compute the marginal posterior state probability $\gamma_t$ and the joint posterior transition probability $\eta_t$:
\begin{subequations}\label{eq:E_step}
\begin{align}
\gamma_t(m,p) &= P(s_t = (m,p) \mid \boldsymbol{b}_{1:\tau}, \lambda), \\
\eta_t((m,p), (j,q)) &= P(s_t = (m,p), s_{t+1} = (j,q) \mid \boldsymbol{b}_{1:\tau}, \lambda),
\end{align}
\end{subequations}
where $s_t \in \mathcal{S}$ denotes the joint agent-mode state at time step $t$ within the observation window.

M-step: Update the specific matrix entries $\hat{T}_{pq}^{mj}$ (which populate the intra-agent blocks $\hat{P}_{mm}$ and cross-agent blocks $\hat{P}_{mj}$) by marginalizing over the observation window length $\tau$:
\begin{equation}\label{eq:M_step}
\hat{T}_{pq}^{mj} = \frac{\sum_{t=1}^{\tau-1} \eta_t((m,p), (j,q))}{\sum_{t=1}^{\tau-1} \gamma_t(m,p)}.
\end{equation}
After each M-step, the block structure of the unnormalized matrix $\tilde{\boldsymbol{T}}$ is updated with these new estimates, scaled by the adjacency weights $w_{jm}$, and the final valid transition matrix $\boldsymbol{T}$ is obtained via the row normalization \eqref{eq:Tnormalization} \cite{simsek2018}.

\subsection{Real-Time Mode Estimation via the Viterbi Algorithm}

Upon receiving a new discretized residual vector $r_m(k)$, it is mapped to $b_m(k)$ and appended to the observation sequence. The Viterbi algorithm~\cite{viterbi1967error, forney1973viterbi} then computes the most likely hidden mode sequence:
\begin{equation}
  \hat{\boldsymbol{\theta}}_{1:\tau} = \argmax_{s_{1:\tau}} P\bigl(\boldsymbol{\theta}_{1:\tau} \mid \boldsymbol{b}_{1:\tau},\, \lambda\bigr),
  \label{eq:viterbi}
\end{equation}
where $\hat{\boldsymbol{\theta}}_k = [\hat{\theta}_1(k), \ldots, \hat{\theta}_M(k)]$ is the estimated joint mode at time $k$.

The complete procedure is summarized in Algorithm~\ref{alg:hmm}.

\begin{algorithm}[t]
\caption{Data-Driven Attack Classification Using HMM}
\label{alg:hmm}
\begin{algorithmic}[1]
\Require Historical sampled residuals $\|r_m(k)\|$ for $m=1,\ldots,M$,\; $k=1,\ldots,\tau$
\State Compute thresholds $\tau^{\text{low}}_{m,\theta}$, $\tau^{\text{high}}_{m,\theta}$ via~\eqref{eq:thresholds}
\State Map each $\|r_m(k)\|$ to bin $b_m(k) \in \{0,1,2,3\}$
\State Form observation vector $\boldsymbol{b}(k)= [b_1(k), b_2(k), \ldots, b_M(k)]$
\State Form observation sequence $\boldsymbol{b}_{1:\tau} = \{\boldsymbol{b}(1),\ldots,\boldsymbol{b}(\tau)\}$
\State Initialize $\lambda = (\boldsymbol{T}, \boldsymbol{O}, \pi)$ via~\eqref{eq:tpm} and~\eqref{eq:opm}
\For{each new time step $k$}
  \State Receive sampled residuals $\|r_m(k)\|$ for all agents $m$
  \State Discretize to obtain $b_m(k)$; append $\boldsymbol{b}(k)$ to $\boldsymbol{b}_{1:k}$
\State Run E-step and M-step to update $\hat{\boldsymbol{T}}$ via \eqref{eq:E_step}, \eqref{eq:M_step}
\State Restore block structure in $\tilde{\boldsymbol{T}}$; renormalize rows to obtain $\boldsymbol{T}$ via \eqref{eq:Tnormalization}  
\State Estimate $\hat{\boldsymbol{\theta}}_k$ via Viterbi algorithm~\eqref{eq:viterbi}
  \State Output mode classification $\hat{\boldsymbol{\theta}}_k$ for all agents
\EndFor
\end{algorithmic}
\end{algorithm}

\section{Simulation and Results}
\label{sec:sim}

\subsection{Network and System Description}

\begin{figure}[t]
  \centering
  \includegraphics[width=0.5\columnwidth]{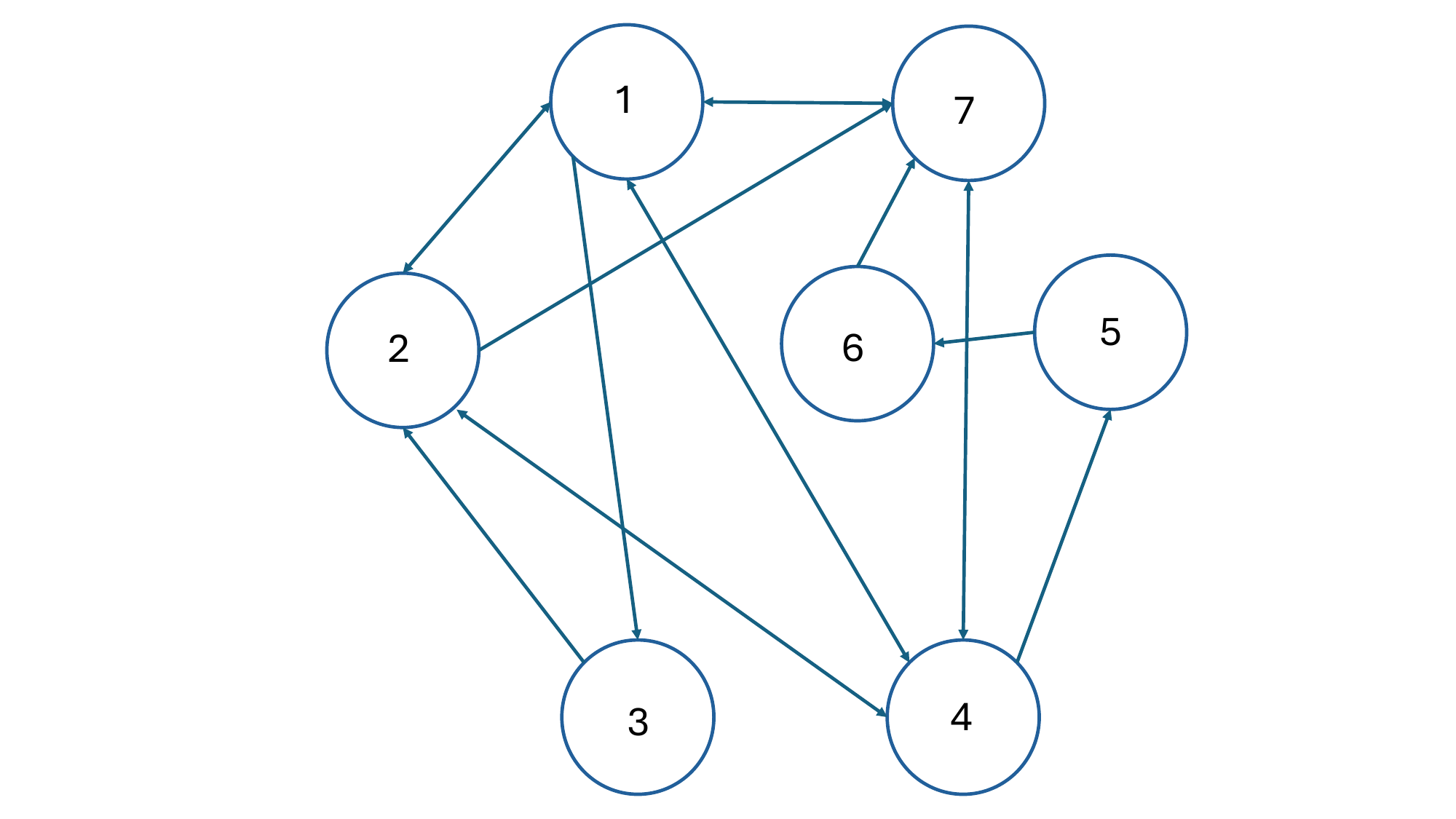}
  \caption{Communication topology of the seven-agent network.}
  \label{fig:topology}
\end{figure}

We consider a MAS of $M = 7$ agents whose directed communication graph is shown in Fig.~\ref{fig:topology}. To generate the full joint observation sequence $\boldsymbol{b}(k)$, each agent is equipped with a local distributed observer that monitors its augmented neighbourhood state. The weighted adjacency matrix is:
\begin{equation}
  W = \begin{bmatrix}
    0 & 3 & 0 & 1 & 0 & 0 & 1\\
    2 & 0 & 1 & 1 & 0 & 0 & 1\\
    1 & 0 & 0 & 0 & 0 & 0 & 0\\
    1 & 1 & 0 & 0 & 0 & 0 & 2\\
    0 & 0 & 0 & 1 & 0 & 0 & 0\\
    0 & 0 & 0 & 0 & 1 & 0 & 0\\
    1 & 1 & 0 & 2 & 0 & 1 & 0
  \end{bmatrix}.
\end{equation}

Each agent in this homogeneous network is modelled as a third-order system with a state vector composed of position, velocity, and angular displacement:
\begin{align*}
  a_m &= \begin{bmatrix}-0.1&1&0\\-0.5&-1.5&0.8\\0&-0.2&-1\end{bmatrix},\quad
  b_m = \begin{bmatrix}0\\0\\0.15\end{bmatrix},\\
  c_m &= \begin{bmatrix}1&0&0\\0&1&0\end{bmatrix},\quad
  f_m = \begin{bmatrix}-0.2&0.01\\-0.1&0.02\\0.1&-0.02\end{bmatrix}, \\
  d_m &= \begin{bmatrix}0.1&0&0\end{bmatrix}^\top.
\end{align*}
The decoy disturbance is modelled as $\xi_m(x_m(t)) = x_{m,1} (t)\cdot\sin(3x_{m,1}(t))$ with $n_\xi = 1$, which is locally Lipschitz continuous within the system's bounded workspace where $L_m = 4$ holds. This signal is injected into the system while it is in decoy mode. FDI profiles span sub-linear growth, sinusoidal, stochastic, ramp, exponential decay, and sigmoid waveforms across the seven agents:
\begin{align*}
\psi_1(t) &= 0.5\sqrt{|t|}, \quad \psi_2(t) = 0.01\sin(t-10), \\
\psi_3(t) &= 0.06\,\mathrm{rand}(t), \quad  \psi_4(t) = -0.01(t-30), \\
\psi_5(t) &= 0.01(t-20),  \quad  \psi_6(t) = 0.9e^{-t}, \\
\psi_7(t) &= 0.6\tanh(t). 
\end{align*}
Modes switch according to a Markov chain with a true intra-agent self-transition probability of $T_{pp}^{mm} = 0.85$ for all modes $p \in \Theta$. The observer residuals are sampled at a period of $h=0.1$~s, yielding one HMM observation per sample. With $T_{pp}^{mm} = 0.85$, the expected dwell time per mode is approximately 7 steps, i.e., $7h = \qty{0.7}{\second}$. The simulation runs for $600$ training steps ($t \in \qty[parse-numbers=false]{[0, 60]}{\second}$), followed by 400 online classification steps ($t \in \qty[parse-numbers=false]{[60, 100]}{\second}$), corresponding to a total duration of \qty{100}{\second}.

\subsection{Threshold Computation}

Mode-dependent thresholds are computed offline from the training residuals via~\eqref{eq:thresholds} with $\kappa_\sigma = 3$ and are summarized in Table~\ref{tab:thresholds}. The spread of the intervals varies across agents, reflecting the heterogeneous FDI signal magnitudes. These bounds define the bin edges used to discretize residuals for HMM-based mode classification.

\begin{table}[t]
  \centering
  \caption{Mode-dependent residual norm threshold ranges per agent ($\kappa_\sigma = 3$). Each entry covers $[\tau^{\text{low}},\,\tau^{\text{high}}]$ across all four modes.}
  \label{tab:thresholds}
  \begin{tabular}{ccc}
    \toprule
    Agent & $\tau^{\text{low}}$ range & $\tau^{\text{high}}$ range \\
    \midrule
    1 & [-4.15,~-1.68] & [2.43,~6.83] \\
    2 & [-2.88,~-1.65] & [2.19,~4.88] \\
    3 & [-3.21,~-1.15] & [1.98,~5.30] \\
    4 & [-3.14,~-2.09] & [2.98,~5.69] \\
    5 & [-3.26,~-2.02] & [2.99,~5.94] \\
    6 & [-2.63,~-1.71] & [2.19,~4.68] \\
    7 & [-3.10,~-1.59] & [2.91,~5.58] \\
    \bottomrule
  \end{tabular}
\end{table}

\subsection{Consensus and State Behaviour}

\begin{figure}[t]
  \centering
  \includegraphics[width=0.8\columnwidth]{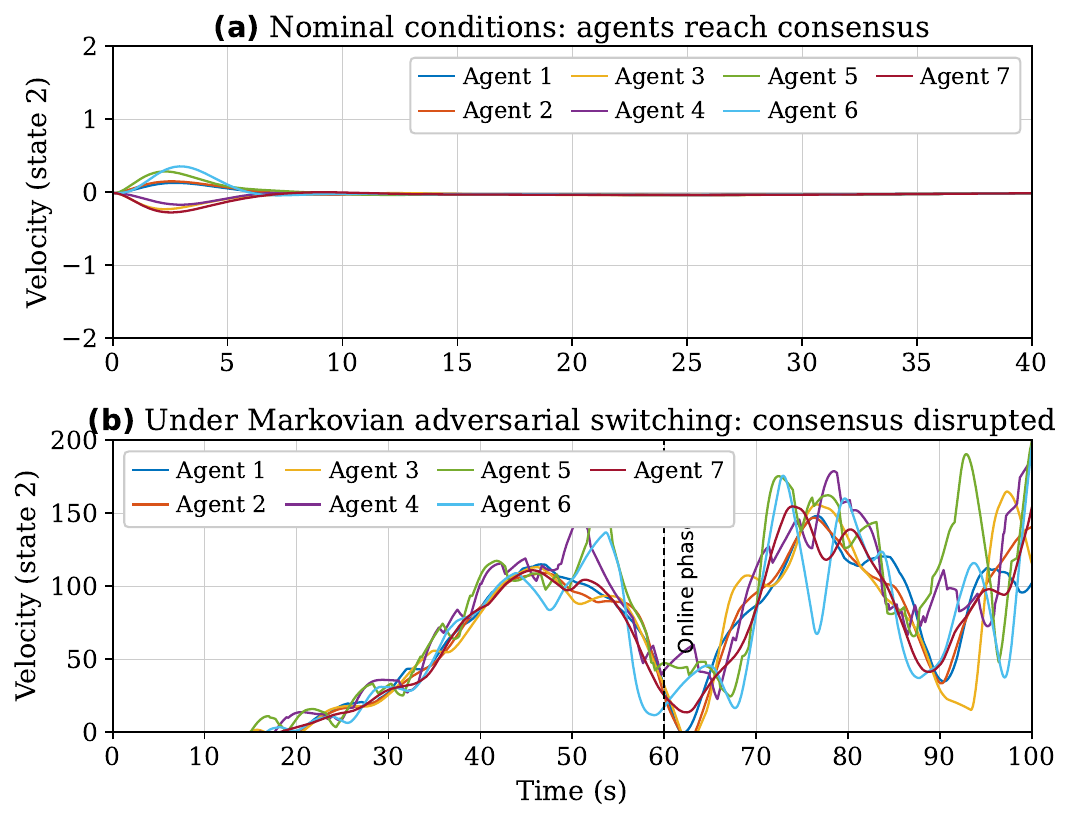}
  \caption{Velocity (second component of the state $x_m$) evolution of all seven agents. (a)~Nominal conditions. (b)~Under Markovian adversarial switching (FDI, DoS, and decoy modes): consensus is disrupted. The dashed vertical line marks the boundary between the training and online phases.}
  \label{fig:consensus}
\end{figure}

Fig.~\ref{fig:consensus} compares the velocity trajectory (state~2) of all seven agents under two scenarios. Under nominal conditions (Fig.~\ref{fig:consensus}(a)), the cooperative control law drives all agents to consensus within a finite time horizon. Once stochastic adversarial mode switching is activated (Fig.~\ref{fig:consensus}(b)), the trajectories diverge and consensus is disrupted, establishing the need for the proposed detection and classification framework.

\subsection{Classification Performance}

\begin{figure}[t]
  \centering
  \includegraphics[width=0.8\columnwidth]{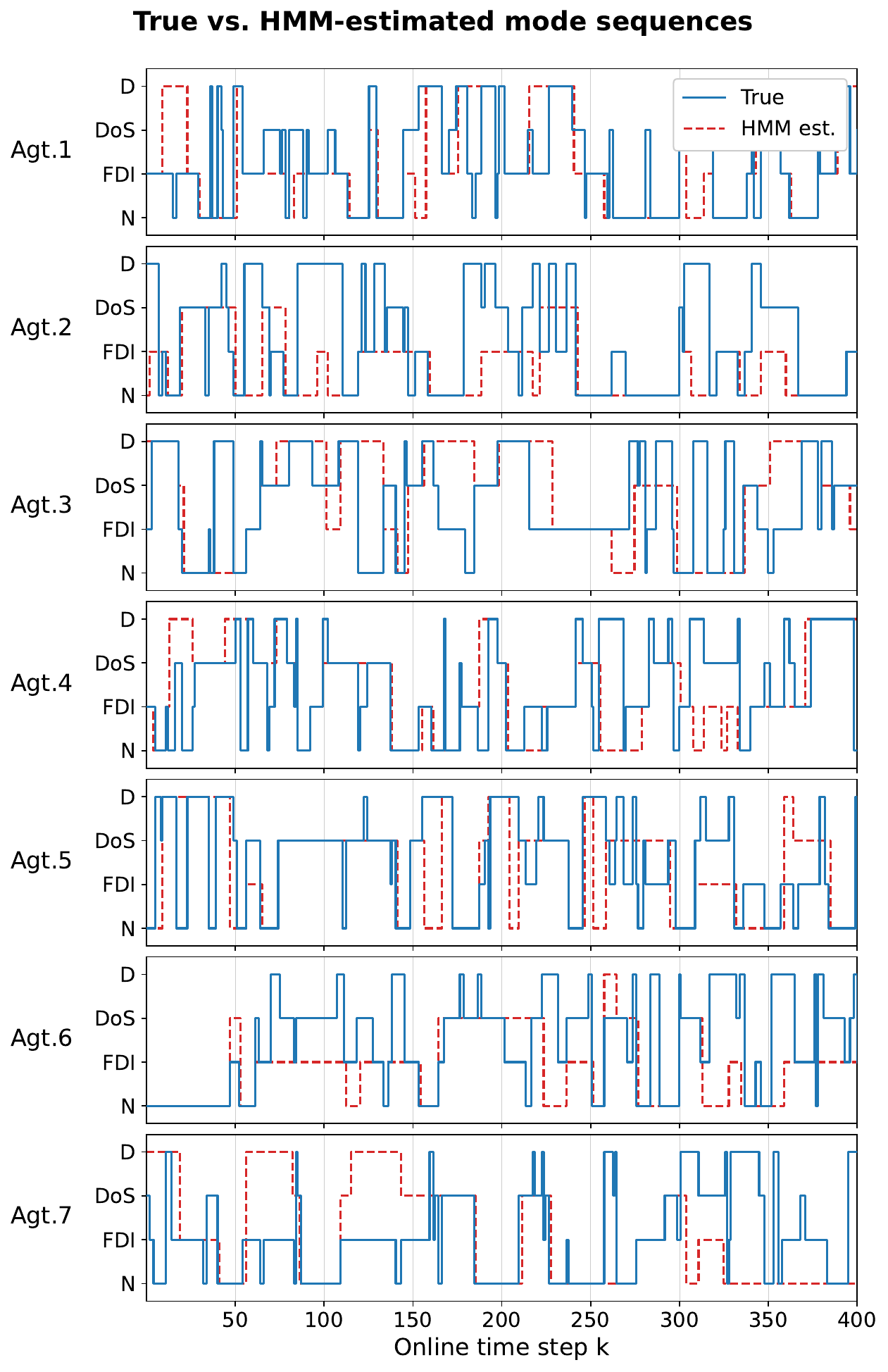}
  \caption{True (blue solid) and HMM-estimated (red dashed) mode sequences for all seven agents over the online classification phase ($400$ time steps). Mode labels: N\,=\,nominal, FDI\,=\,false-data injection, DoS\,=\,denial-of-service, D\,=\,decoy.}
  \label{fig:modes}
\end{figure}

Fig.~\ref{fig:modes} displays the true and HMM-estimated mode sequences for all seven agents over the $400$-step online phase. The joint Viterbi decoder tracks the true mode sequence closely across all agents, with mismatches concentrated at transition instants, i.e., the time steps at which the residual signal carries insufficient evidence to resolve the incoming mode before the next observation arrives. Crucially, the decoy mode (D) is consistently distinguished from the nominal mode (N) despite their overlapping residual distributions, confirming that the graph-structured TPM provides the contextual information needed to resolve this ambiguity.

Table~\ref{tab:accuracy} reports per-mode precision, recall, and F1-score aggregated across all seven agents. The decoy mode yields the lowest but still reliable precision.

\begin{table}[t]
  \centering
  \caption{Per-mode classification metrics aggregated across all seven agents ($400$ steps).}
  \label{tab:accuracy}
  \begin{tabular}{lcccc}
    \toprule
    Metric & Nominal & FDI & DoS & Decoy \\
    \midrule
    Precision (\%) & 63.5 & 47.1 & 55.1 & 46.7 \\
    Recall    (\%) & 81.2 & 44.5 & 51.0 & 36.3 \\
    F1-score  (\%) & 71.3 & 45.8 & 52.9 & 40.9 \\
    \bottomrule
  \end{tabular}
\end{table}

\begin{figure}[t]
  \centering
  \includegraphics[width=0.82\columnwidth]{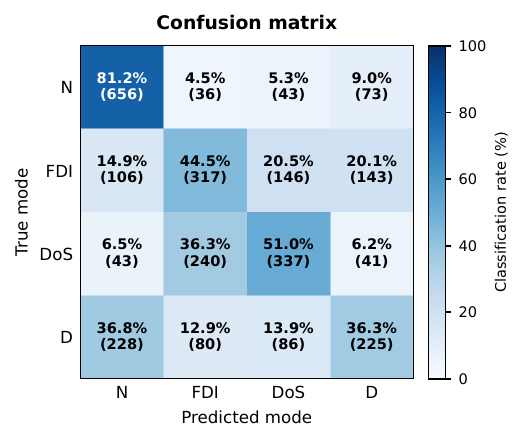}
  \caption{Normalized confusion matrix pooled across all seven agents (online phase). Each cell shows the classification rate (\%) and raw count. Diagonal dominance confirms reliable four-mode classification; residual off-diagonal mass between Decoy and Nominal reflects their overlapping residual distributions.}
  \label{fig:confusion}
\end{figure}

Fig.~\ref{fig:confusion} shows the normalized confusion matrix pooled across all agents. The diagonal dominance confirms reliable classification of all four modes. Off-diagonal entries are largest between Decoy and Nominal, reflecting the inherent statistical overlap engineered into the decoy signal, a challenge that the structured OPM and EM update jointly address by leveraging cross-agent residual context.

\begin{figure}[t]
  \centering
  \includegraphics[width=\columnwidth]{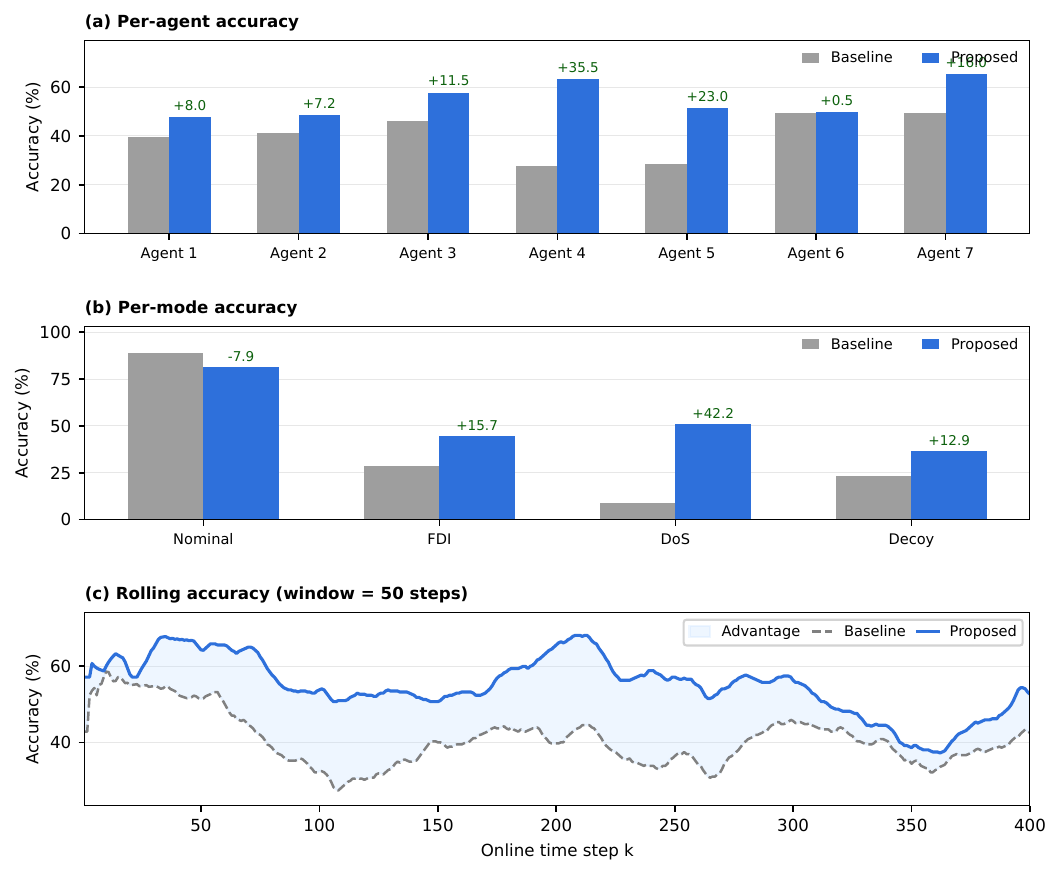}
  \caption{Three-panel comparison of the proposed joint 28-state HMM against the per-agent baseline (no topology coupling). \textbf{(a)}~Per-agent accuracy; numbers above bars indicate gain in percentage points. \textbf{(b)}~Per-mode accuracy. \textbf{(c)}~Rolling accuracy (window\,=\,50 steps).}
  \label{fig:comparison}
\end{figure}

To isolate the contribution of the topology-coupled joint HMM, we compare the proposed framework against a baseline in which each agent is classified independently using only its own $4 \times 4$ diagonal TPM block, exactly the approach used in prior per-agent Viterbi schemes that ignore cross-agent coupling~\cite{viterbi1967error, forney1973viterbi}. Fig.~\ref{fig:comparison} summarizes the comparison across three views: per-agent accuracy, per-mode accuracy, and rolling accuracy over time.

The proposed joint decoder consistently outperforms the baseline across all agents and all modes, confirming that inter-agent context, encoded in the off-diagonal TPM blocks scaled by the adjacency weights, is the decisive factor in resolving mode classification ambiguity. The rolling accuracy plot further shows that the advantage is sustained throughout the online phase rather than being concentrated at a particular time window.

\section{Conclusion}
\label{sec:conclusion}

This paper presents a topology-coupled Hidden Markov Model framework for online detection and classification of multi-modal adversarial threats in networked multi-agent systems. By decoupling continuous time unknown false-data injection inputs via distributed observers and mapping the resulting residuals to a graph-structured transition probability matrix, the proposed method reliably distinguishes Lipschitz-bounded decoy perturbations, designed to evade detection, from benign operations. Furthermore, integrating an online Expectation-Maximization update allows the classifier to adapt to shifting adversarial modes without prior knowledge of the attacker's operational policy. Numerical simulations on a directed seven-agent network validated that incorporating inter-agent communication topology into the joint transition dynamics significantly improves classification precision and recall over decoupled, agent-independent estimation schemes, leveraging the local communication topology to detect deceptive attacks.

Several possible directions are left for future research. First, extending the transition model to evolving topologies subject to stochastic link failures would enable attack tracking under dynamic communication degradation. Second, replacing the Gaussian emission likelihoods with semi-parametric or neural representations could improve classification fidelity when residuals deviate significantly from unimodal distributions under aggressive non-stationary interference. Third, extending the current local observer architecture to a sparse, topology-aware deployment strategy, such as placing observers selectively at strongly connected components~\cite{dorogovtsev2001} to minimize sensing and communication overhead in large-scale networks, represents a valuable direction for future study. Finally, establishing analytical bounds on classification error as a function of the graph spectral gap, observer sampling period $h$, and EM convergence rate remains an important theoretical objective.

\appendix
\section{Unknown-Input Observer Design}
\label{app:uio}

Adversarial FDI signals enter the system as unknown inputs whose magnitude and form are inaccessible to the defender. A UIO is therefore chosen as the estimator to decouple the FDI contribution from the state estimate, keeping the residuals sensitive to mode changes rather than to the unknown attack signal itself~\cite{barzegar2023}.

For the augmented continuous time system~\eqref{eq:aug_dyn}, the UIO propagates an internal state $R_m(t)$ according to:
\begin{align}
  \dot{R}_m(t) &= G_m R_m(t) + H_m U_m(t) + (I - v_m C_m)D_m \Xi_m \left( \hat{X}_m(t) \right) \nonumber \\
               &\quad + l_m Y_m(t) + Q_m Z_m \bigl(Y_m(t) - \hat{Y}_m(t)\bigr),
  \label{eq:uio1}
\end{align}
and reconstructs the augmented state as:
\begin{equation}
  \hat{X}_m(t) = R_m(t) + v_m Y_m(t). \label{eq:uio2}
\end{equation}
The gain matrix $Z_m$, defined over the augmented set $m \cup \mathcal{N}_m$, encodes the communication topology:
\begin{equation}
  Z_m = \diag\{1, w_{j_1 m}, \ldots, w_{j_{|\mathcal{N}_m|} m}\} \otimes I_{n_y}
\end{equation}
so that neighbours with stronger links contribute more to the correction term. The predicted output is $\hat{Y}_m(t) = C_m \hat{X}_m(t)$, and the observer gain matrices $\{G_m, H_m, v_m, l_m, Q_m\}$ are synthesized via linear matrix inequalities to guarantee the stability of the error dynamics in the presence of the Lipschitz nonlinearity in~\eqref{eq:lipschitz}~\cite{barzegar2023}.

\section*{Acknowledgment}
The authors would like to thank our collaborator Dr.~Bao Nguyen from Defence Research and Development Canada -- Centre for Operational Research and Analysis (DRDC-CORA) for insightful discussions and valuable feedback on this work.

\bibliographystyle{IEEEtran}
\bibliography{mybib}

\end{document}